\documentclass[preprint,12pt]{elsarticle}

\newcommand{\usepackageifexists}[1]{%
     \IfFileExists{#1.sty}{\usepackage{#1}}%
        {\GenericInfo{taglia}{Il package #1 non esiste.}}}

\makeatletter
\newcommand\footnoteref[1]{\protected@xdef\@thefnmark{\ref{#1}}\@footnotemark}
\makeatother

\def\ps@pprintTitle{%
 \let\@oddhead\@empty
 \let\@evenhead\@empty
 \def\@oddfoot{}%
 \let\@evenfoot\@oddfoot}

\usepackage{verbatim}
\usepackage{amsmath,amsthm,amssymb}
\usepackage{upref}
\usepackage{graphicx, caption, subcaption}

\usepackage{listings}
\usepackage{epstopdf}

\usepackage{color}
\usepackage{mathrsfs}
\usepackage{multirow}
\providecommand{\mathscr}[1]{\mathcal{#1}}

\usepackage{graphicx}
\usepackageifexists{pdfsync}
\graphicspath{ {figures/} }
\usepackage{booktabs}
\usepackage{hyperref}
\usepackage{geometry, natbib} 
\usepackage{adjustbox}
\usepackage[autostyle]{csquotes} 
\usepackage{enumerate}
\usepackage[flushleft]{threeparttable}

\def\correspondingauthor{\footnote{Corresponding author.}}

\begin{document}
\def\spacingset#1{\renewcommand{\baselinestretch}%
	{#1}\small\normalsize} \spacingset{1}

\title{On The Calibration of Short-Term Interest Rates Through a CIR Model}

\author[Orlando]{Giuseppe Orlando\correspondingauthor{}}
\address[Orlando]{Dipartimento di Scienze Economiche e Metodi Matematici, Universit\`a degli Studi di Bari Aldo Moro, Italy}

\author[Mininni]{Rosa Maria Mininni\footnote{\label{note1}The contribution of both authors to this work is equivalent.}} 
\address[Mininni]{Dipartimento di Matematica, Universit\`a degli Studi di Bari Aldo Moro,Italy}
 
\author[Bufalo]{Michele Bufalo\footnoteref{note1}} 
\address[Bufalo]{Dipartmento di Metodi e Modelli  per l'Economia, il Territorio e la Finanza, Universit\`a degli Studi di Roma ``La Sapienza", Italy} 

\begin{abstract}
It is well known that the Cox-Ingersoll-Ross (CIR) stochastic model to study the term structure of interest rates, as introduced in 1985, is inadequate for modelling the current market environment with negative short interest rates. Moreover, the diffusion term in the rate dynamics goes to zero when short rates are small; both volatility and long-run mean do not change with time; they do not fit with the skewed (fat tails) distribution of the interest rates, etc. The aim of the present work is to suggest a new framework, which we call  the \emph{CIR\# model}, that well fits the term structure of short interest rates so that the market volatility structure is preserved as well as the analytical tractability of the original CIR model.

\end{abstract}

\begin{keyword} Interest rates forecasting, volatility, ARIMA models, simulation, jumps fitting, translation\\
JEL Classification: G12, E43, E47
\MSC[2010] 91G30, 91B84, 91G60, 91G70, 62M10
\end{keyword}

\maketitle

\section{Introduction}\label{section1}

The aim of the present work is to provide a new numerical methodology for the CIR framework, which we call  the \emph{CIR\# model}, that well fits the term structure of  short interest rates as observed in a real market. Our approach is based on a proper translation of interest rates so that the market volatility structure is preserved as well as the analytical tractability of the original CIR model. 

Cox, Ingersoll \& Ross (1985) \cite{CIR_85} proposed a term structure model, well known as the CIR model, to describe the price of discount zero-coupon bonds with variuous maturities under no-arbitrage condition. This model generalizes the Vasicek model \cite{Vasicek1977} to the case of non constant volatility and assumes that the evolution of the underlying short term interest rate is a diffusion process, i.e. a continuous Markov process unique solution to the following stochastic differential equation (SDE)
\begin{equation}\label{CIRrate}
dr(t)  = [k(\theta-r(t)) - \lambda(t,r(t))]dt + \sigma \sqrt{r(t)}dW(t),
\end{equation}
with initial condition $r(0)= r_{0} >0 $. $(W(t))_{t\ge 0} $ denotes a standard Brownian motion  under the risk neutral probability measure, intended to model a random risk factor. The interest rate process $(r(t))_{t\ge 0}$ is usually known as the {\it CIR process} or {\it square root process}.
The SDE \eqref{CIRrate} is classified as a one-factor time-homogeneous model, because the parameters $k, \theta$ and $\sigma, $ are time-independent and the short interest rate dynamic is driven only by the market price of risk $ \lambda(t,r(t)) := \lambda r(t), $ where $\lambda$ is a constant. Therefore the SDE \eqref{CIRrate} is composed of  two parts: the ``mean reverting" drift component $k[\theta-r(t)],$ which ensures the rate $r(t)$ is elastically pulled towards a long-run mean value $\theta >0$ at a speed of adjustment $k>0,$ and the random component $W(t),$ which is scaled by the standard deviation $\sigma \sqrt{r(t)}$. The volatility of the instantaneous short rate is denoted by $\sigma>0$.  

The paths of the CIR process never reach negative values and their behaviour depends on the relationship between the three constant positive parameters $k, \theta, \sigma$.  Indeed, it can be shown (see, e.g., Jeanblanc, Yor \& Chesney (2009) \cite[Ch. 3]{Jeanblanc2009}) that if  the condition $2k\theta \ge\sigma ^{2}$ is satisfied, then the interest rates
$r(t)$ are strictly positive for all $t>0$, and, for small $r(t)$, the process rebounds as the random perturbation dampens with $r(t)\rightarrow 0$.   
Furthermore, the CIR process belongs to the class of processes satisfying the ``affine property", i.e.,  the logarithm of the characteristic function of the transition distribution of such processes is an affine (linear plus constant) function with respect to their initial state (for more details  the reader can refer to Duffie, Filipovi{\'c} \& Schachermayer (2003) \cite[Section 2]{Duffie_Filipovic}. As a consequence, the non-arbitrage price of a discount zero coupon bond with maturity $T>0$ and  underlying interest rate dynamics  described by a CIR process, is given by
\begin{equation}\label{CIRprice}
P(T-t,r(t)) = A(T-t)e^{ -B(T-t)r(t)},\quad  t\in [0,T],
\end{equation}
where $A(\cdot)$ and $B(\cdot)$ are deterministic functions (see Cox, Ingersoll \& Ross (1985) \cite{CIR_85}). The final condition is $P(T,r(T))  = 1,$ which corresponds to the nominal value of the bond conventionally set equal to 1 (monetary unit). Since bonds are commonly quoted in terms of yields rather than prices, the formula \eqref{CIRprice} allows derivation of the yield-to-maturity curve
\begin{align*}
Y(T-t,r(t)) &:= - \ln{P(t,T,r(t))}/(T-t) \\[2.\jot]
&= [B(T-t)r(t) - \ln(A(T-t))]/(T-t),\quad  t\in [0,T],
\end{align*}
which tends to the asymptotic value $\varUpsilon=2k\theta/(\gamma + k + \lambda)$  as $T\to\infty, $ where $\gamma := \sqrt{(k+\lambda)^2 + 2\sigma^2}.$  

Thus the CIR process is characterized by the following properties: 1) short interest rates never become negative; 2) if the interest rate reaches the zero value (it may occur under the condition $2k\theta <\sigma ^{2}$), it is immediately reflected in positive values; 3) the diffusion term in \eqref{CIRrate} increases when $r(t)$ increases; 4) the transition density of short rates is a noncentral chi-squared distribution, which converges to a gamma distribution as time grows towards infinity; 5) the trajectories describing the short rate dynamics cannot be explicitly derived, but exact simulation is still possible and bond prices are explicitly computable from it.


The remainder of the paper is organized as follows. Section 2 summarizes the existing literature on the CIR model and the related extensions. Section 3 describes the principal steps of the proposed \emph{CIR\# model}; Section 4 presents in more detail the numerical procedure and tests the goodness-of-fit of the new methodology to market data. Finally, Section 5 concludes.

\section{Literature review}
The CIR model became very popular in finance among practitioners because it was perceived as an improvement on the Vasicek model, not allowing for negative rates and introducing a rate dependent volatility, as well as for its relatively handy implementation and analytical tractability. 
Other applications include stochastic volatility modelling in option pricing problems (see Orlando and Taglialatela (2017)  \cite{Orlando_T}, Heston (1993) \cite{Heston}), or  default intensities in credit risk (see Duffie (2005) \cite{DuffieC}).

\begin{figure}[ht!]
	\centering
	\includegraphics[width=8cm]{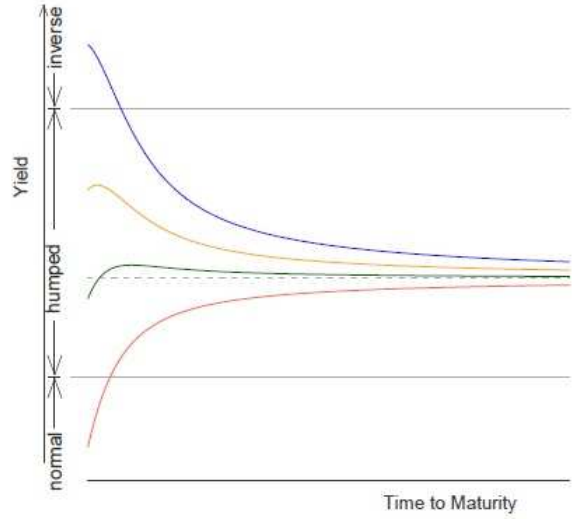}
	\caption{Inverse, humped or normal yield-to-maturity curves (Keller-Ressel \& Steiner(2008) \cite[Figure 1]{Keller}).}
	\label{Fig-KRS}
\end{figure}

Despite the previously listed properties, the CIR model fails as a satisfactory calibration to market data  since it depends on a small number of constant parameters, $k,\theta$ and $\sigma.$  
As proved by Keller-Ressel and Steiner (2008) \cite[Theorem 3.9 and Section 4.2]{Keller} the yield-to-maturity curve of any time-homogenous, affine one-factor model is either normal (i.e., a strictly increasing function of $T-t$), humped (i.e., with one local maximum and no minimum on $]0,\infty[$) or inverse (i.e., a strictly decreasing function of $T-t$) (see Figure \ref{Fig-KRS}).
For the CIR process, the yield is normal when $r(t) \le k\theta/(\gamma - 2(k + \lambda))$, while it is inverse when $r(t) \ge k\theta/(k + \lambda).$ For intermediate values the yield curve is humped.  

\begin{figure}[ht!]
\centering    
\begin{subfigure}[d]{0.65\textwidth}
\includegraphics[width=1.2\textwidth]{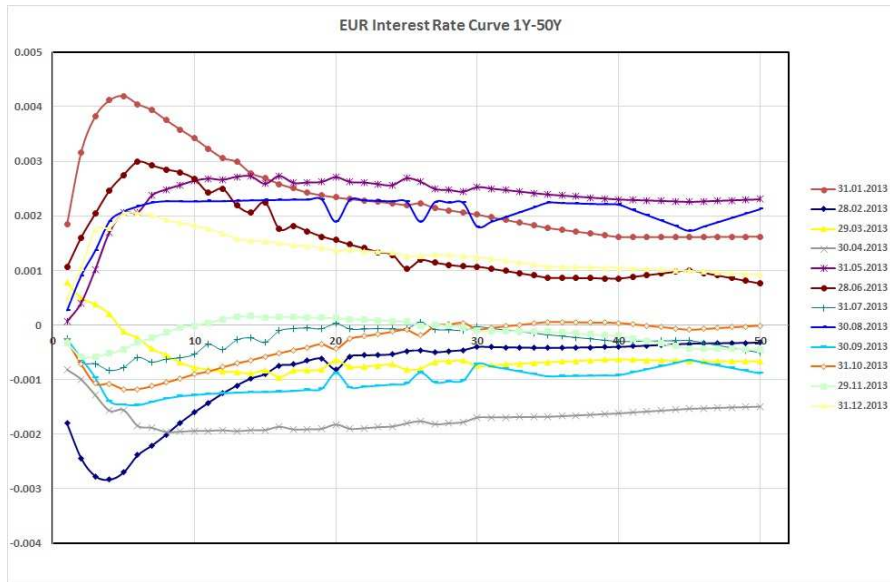}
\caption{\scriptsize EUR term structure through 2013.}
\label{fig:mouse}
\end{subfigure}
\begin{subfigure}[e]{0.65\textwidth}
	\includegraphics[width=1.2\textwidth]{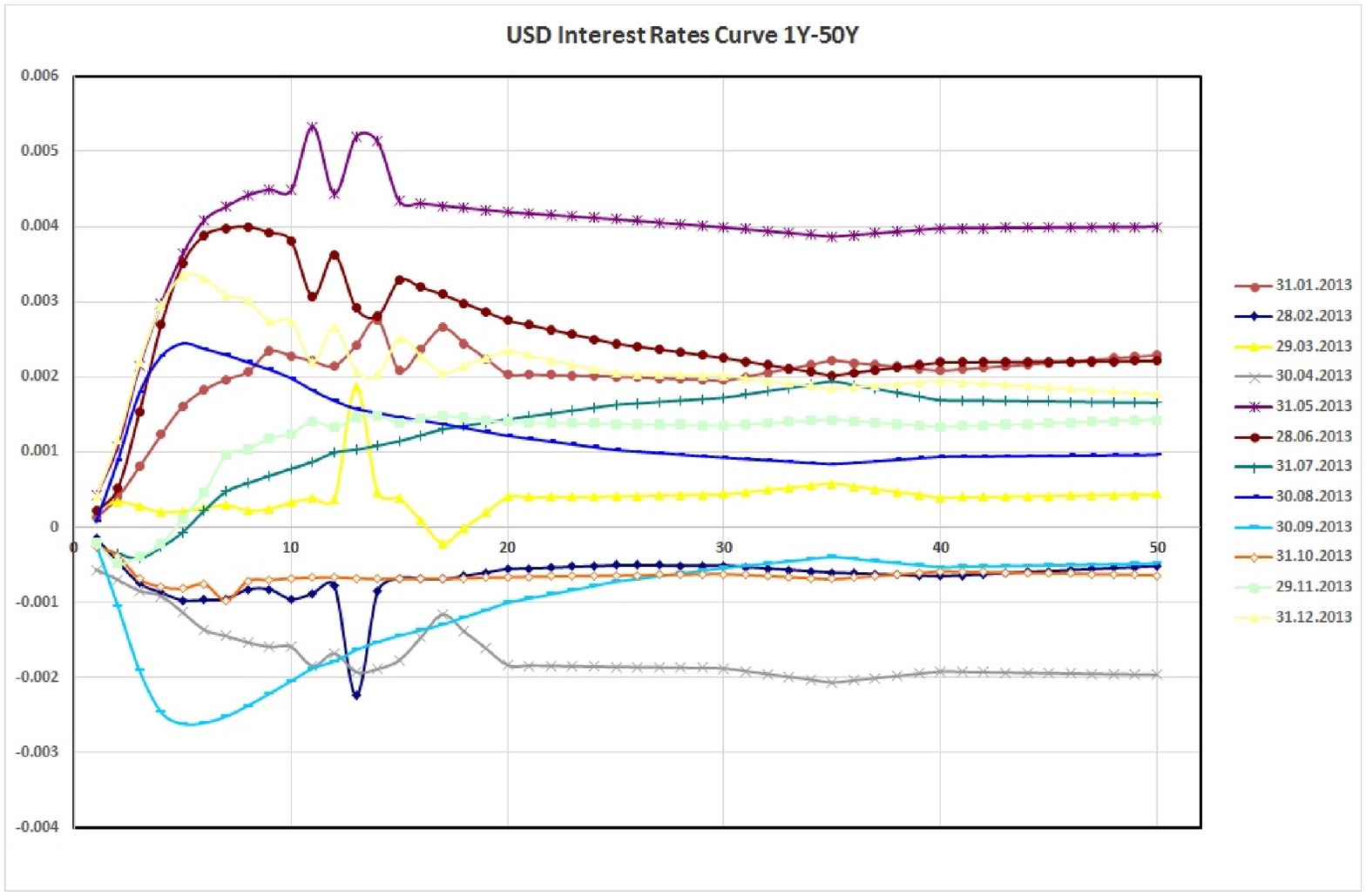}
  \caption{{\scriptsize  USD term structure through 2013.}}
\label{fig:mouse}
\end{subfigure}
\caption{EUR and USD term structure (1Y-50Y) as observed on monthly basis from January to December 2013.}\label{fig:EUR_USD-Term_Str}
\end{figure}

However, Carmona and Tehranchi (2006), \cite[Section 2.3.5]{Carmona} explained that: ``Tweaking the parameters can produce yield curves with one hump or one dip (a local minimum), but it is very difficult (if not impossible) to calibrate the parameters so that the hump/dip sits where desired. There are not enough parameters to calibrate the models to account for observed features contained in the prices quoted on the markets" (see Figure \ref{fig:EUR_USD-Term_Str}). 

Thus the need for more sophisticated models for an exact fit to the currently-observed yield curve,  which could take into account multiple correlated sources of risk  as well as shocks and/or structural changes of the market, led some years later to the development of extensions of the CIR model. Among the best known we mention: the Hull-White (1990) \cite{Hull1990} model based on the idea of considering time-dependent coefficients; the Chen (1996) \cite{Chen} three-factor model; the CIR++ model by Brigo \& Mercurio (2001) \cite{Brigo2001}  that considers short rates shifted by a deterministic function chosen to fit exactly the initial term structure of interest rates; the jump diffusion JCIR model (see Brigo \& Mercurio (2006) \cite{BrigoMercurio2006}) and JCIR++ by Brigo \& El-Bachir (2006)\cite{Brigo2006} where jumps are described by  a time-homogeneous Poisson process;  the CIR2 and CIR2++ two-factor models (see  Brigo \& Mercurio (2006)  \cite{BrigoMercurio2006}). Very recently, Zhu (2014) \cite{Zhu}, in order to incorporate the default clustering effects, proposed a  CIR process with jumps modelled by a Hawkes process (which is a point process that has self-exciting property and the desired clustering effect), Moreno et al. (2015) \cite{Moreno} presented a cyclical square-root model,  and Najafi et al. (2017) (\cite{Najafi1}, \cite{Najafi2})  proposed some extensions of the CIR model where a mixed fractional Brownian motion applies to display the random part of the model. 

\begin{figure}[ht!]
\centering
	\includegraphics[width=12cm]{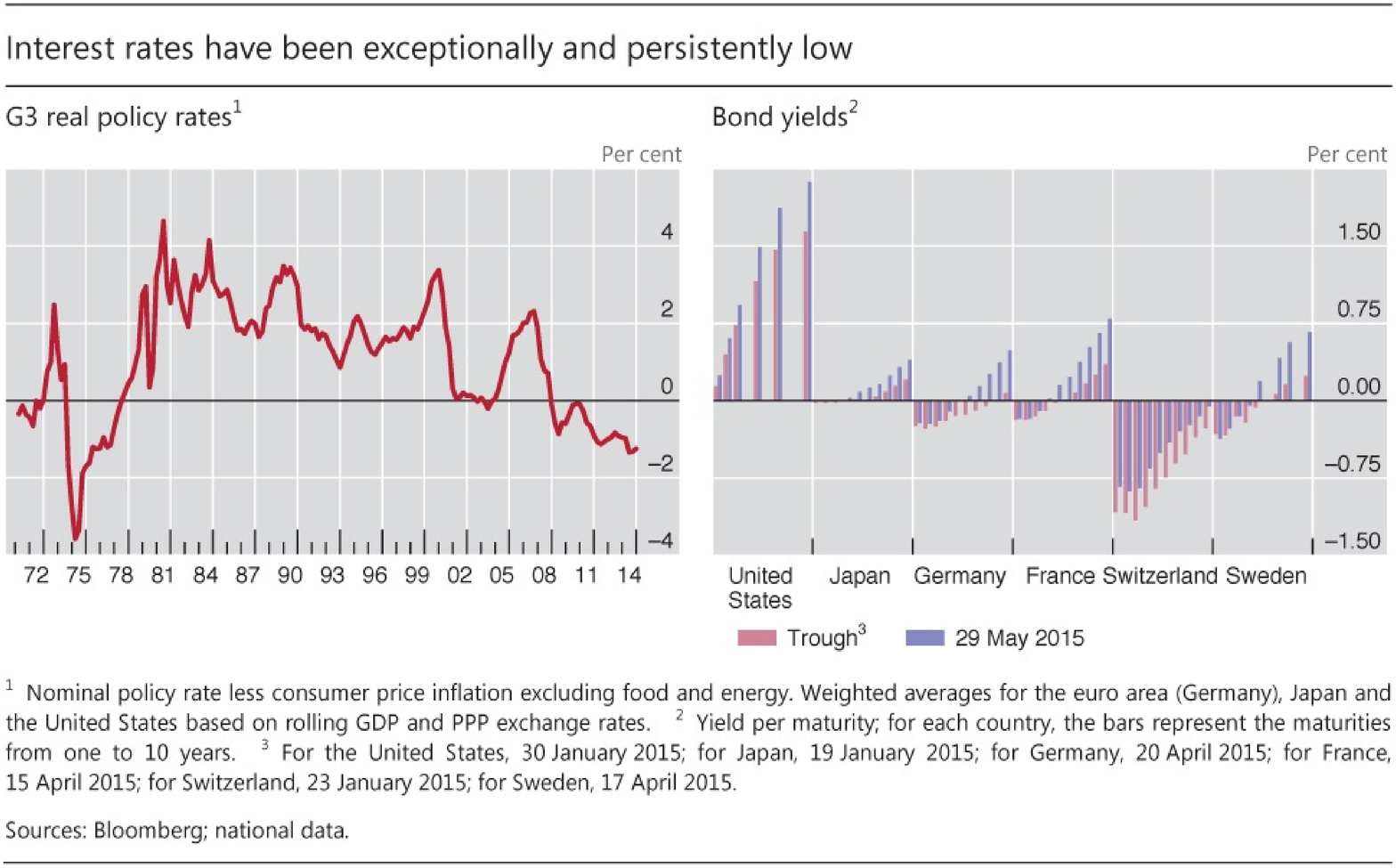}
	\caption{$ $BIS 85th Annual Report 2015.}
	\label{Fig-BIS-2015-G1}
\end{figure}

Note that all the  above cited extensions to CIR model preserve the positivity of interest rates, in some cases through reasonable restrictions on the parameters. But the financial crisis of 2008 and the ensuing quantitative easing policies brought down interest rates, as a consequence of reduced growth of developed economies, and accustomed markets to unprecedented negative interest regimes under the so called ``new normal". As observed in Engelen (2015) \cite{Engelen} and BIS (2015) \cite{BIS}: 

\noindent ``Interest rates have been extraordinarily low for an exceptionally long time, in nominal and inflation-adjusted terms, against any benchmark"(see Figure \ref{Fig-BIS-2015-G1}). ``Between December 2014 and end-May 2015, on average around \$2 trillion in global long-term sovereign debt, much of it issued by euro area sovereigns, was trading at negative yields", ``such yields are unprecedented. Policy rates are even lower than at the peak of the Great Financial Crisis in both nominal and real terms. And in real terms they have now been negative for even longer than during the Great Inflation of the 1970s. Yet, exceptional as this situation may be, many expect it to continue". "Such low rates are the most remarkable symptom of a broader malaise in the global economy: the economic expansion is unbalanced, debt burdens and financial risks are still too high, productivity growth too low, and the room for manoeuvre in macroeconomic policy too limited. The unthinkable risks becoming routine and being perceived as the new normal."

Therefore, the need for adjusting  short term interest rate models for negative rates has become an additional characteristic that a ``good" model should possess. It is worth noting that the main drawback of the Vasicek model  \cite{Vasicek1977}, which allows for negative interest rates,  is that the conditional volatility of changes in the interest rate is constant, independent on the level of it, and this may unrealistically affect the prices of bonds that can grow exponentially (see Rogers \cite{Rogers}). For this reason, the Vasicek model is unused by practitioners.

\section{The CIR\# model}\label{section2}

In the following we will illustrate our original approach, but first let us recap the main issues of the CIR model: 
\begin{enumerate}[\bf i.]
\item Negative interest rates are precluded; \label{item1}
\item The diffusion term in \eqref{CIRrate} goes to zero when $r(t)$ is small (in contrast with market data); \label{item2}
\item The instantaneous volatility $\sigma$ is constant (in real life $\sigma$ is calibrated continuously from market data);\label{item3}
\item There are no jumps (e.g. caused by government fiscal and monetary policies, by release of corporate financial results, etc.);\label{item4}
\item Risk premia are linear with interest rates (false if credit worthiness of a counterparty and market volatility are considered);  \label{item5}
\item The change in interest rates depends only on the market risk. \label{item6}
\end{enumerate}

The aim of the present work is to provide a new methodology that gives an answer to points {\bf i.- iv.} by preserving the structure of the original CIR model \eqref{CIRrate} to describe the dynamics of spot interest rates observed in financial markets. For this purpose the first step is partitioning the available market data sample into sub-samples - not necessarily of the same size - in order to capture all the statistically significant changes of variance in real spot rates and consequently, to give an account of jumps (see Section \ref{section3.1}). This should allow to overcome the critical issue pointed out in {\bf iv.}. After that, to overcome challenges {\bf i.- ii.}, the real spot rates are properly translated to shift them  away from zero or negative values and such that the diffusion term in \eqref{CIRrate} is not dampened by the proximity to zero but fully reflects the same level of volatility present on the market (see Section \ref{section3.2}). 

The second step consists in fitting an ``optimal" - as explained in Section \ref{section3.4} - ARIMA model to each sub-sample of market data.  To ensure that the residuals of the chosen ``optimal" ARIMA model in each sub-sample look like Gaussian white noise, the Johnson's transformation (Johnson (1949)\cite{Johnson1949}) is applied to the standardized residuals (see Section \ref{section3.3}).

As a third step, the parameters $k, \theta, \sigma$ in \eqref{CIRrate} are calibrated to the (eventually) shifted market interest rates by estimating them for each sub-sample of available data, as explained in Section \ref{section3.3.1} (which allows to overcome the issue {\bf iii.}). For this purpose, trajectories of the CIR process are simulated by a strong convergent discretization scheme, using the standardized residuals of the ``optimal" ARIMA model selected for each sub-sample in place of realizations of a standard Brownian motion. As a result, exact CIR fitted values to real data are calculated and the computational cost of the numerical procedure is considerably reduced. Finally, the short-term interest rates estimated by the CIR model are shifted back and compared to real data. As a measure of goodness-of-fit to the available market data,  we compute:
\begin{itemize}
	\item the statistics $R^2$ given by the following expression \cite{Kvalseth}
\begin{equation}\label{rsquare}
R^2 = 1 - \frac{\sum\limits_{h=1}^{m} (e_h - \overline{e})^2}{\sum\limits_{h=1}^{m} (r_h - \overline{r})^2},
\end{equation}
where $e_h = r_h - \widehat{r}_h$ denotes the residual between the observed market interest rate $r_{h}$ and the corresponding fitted value $\widehat{r}_h,$ evaluated on a data sample of size $m\ge 2.$  Furthermore, $\overline{e}$ and $\overline{r}$ denote the sample mean of $e_h$ and $r_{h}$,  respectively; 
 
\item the square root of the mean square error (RMSE)
\begin{equation}
\varepsilon=\sqrt{\frac{1}{m}\sum_{h=1}^{m} {e^2_h}}.
\end{equation}
\end{itemize}

\section{Numerical Implementation and Empirical Analysis}\label{section3}

\subsection{The Dataset}\label{sec:Dataset}
\noindent In this section we give explicit numerical results for the \emph{CIR\# model} described in Section \ref{section2}. 
Our dataset records EUR and USD interest rates with maturities 1/360A--360/360A and 1Y--50Y (i.e. at 1 day (overnight), 30 days, 60 days,...., 360 days  and  1 Year,...,50 Years) available from  IBA \cite{IBA}. For each maturity, interest rates are recorded on monthly basis from 31 December 2010 to 29 July 2016 for a total number of 68 datapoints. In the book \cite{OMB1}(2018) we performed a qualitative analysis on this dataset.
We found that the most challenging task was to fit short-term interest rates with maturity 1/360A, 30/360A, 60/360A, 90/360A, 120/360A,..., 360/360A, respectively, due to the presence of next-to-zero and/or negative spot rate values. This led us to implement a novel numerical procedure, the  CIR$\#$ model, that would allow description of the short-term structure by the original CIR model. The new methodology will be discussed in more details in the next subsections and we will give explicit numerical results for the CIR$\#$ model applied to the data sample consisted of $n=68$ EUR interest rates with 1 day (overnight) maturity. All computations have been executed using MATLAB\textsuperscript{\textregistered} R$2017$a.

\subsection{ANOVA test with a fixed segmentation}\label{section3.1}
As explained in Section \ref{section2}, our first objective is to overcome the issues pointed out in {\bf i., ii.} and {\bf iv.} for the CIR model. Thus we start to partition the whole data sample into sub-samples, which we call {\em groups}, by a one-way ANOVA analysis to highlight statistically significant changes of variance in real spot rates and so to give an account of possible jumps. The main difficulty concerns the choice of the optimal partition into groups to apply the ANOVA test; we had to take into account both the size (the smaller the group is, the more refined the analysis) and the ability to capture any jumps (the larger the group, the better in terms of statistical significance).

After several tests, we decided  to segment the whole sample into eight groups each of size $m=8$ or a multiple thereof (except for the last group, obviously). The results of the one-way ANOVA test are reported in Table \ref{tab:ANOVA}. The \textit{p-value} (Prob$>$F) of $8.00796\cdot 10^{-19}$ indicates a statistically significant difference between groups.

\begin{figure}[h]
	\centering
	{\includegraphics[width=0.8\textwidth]{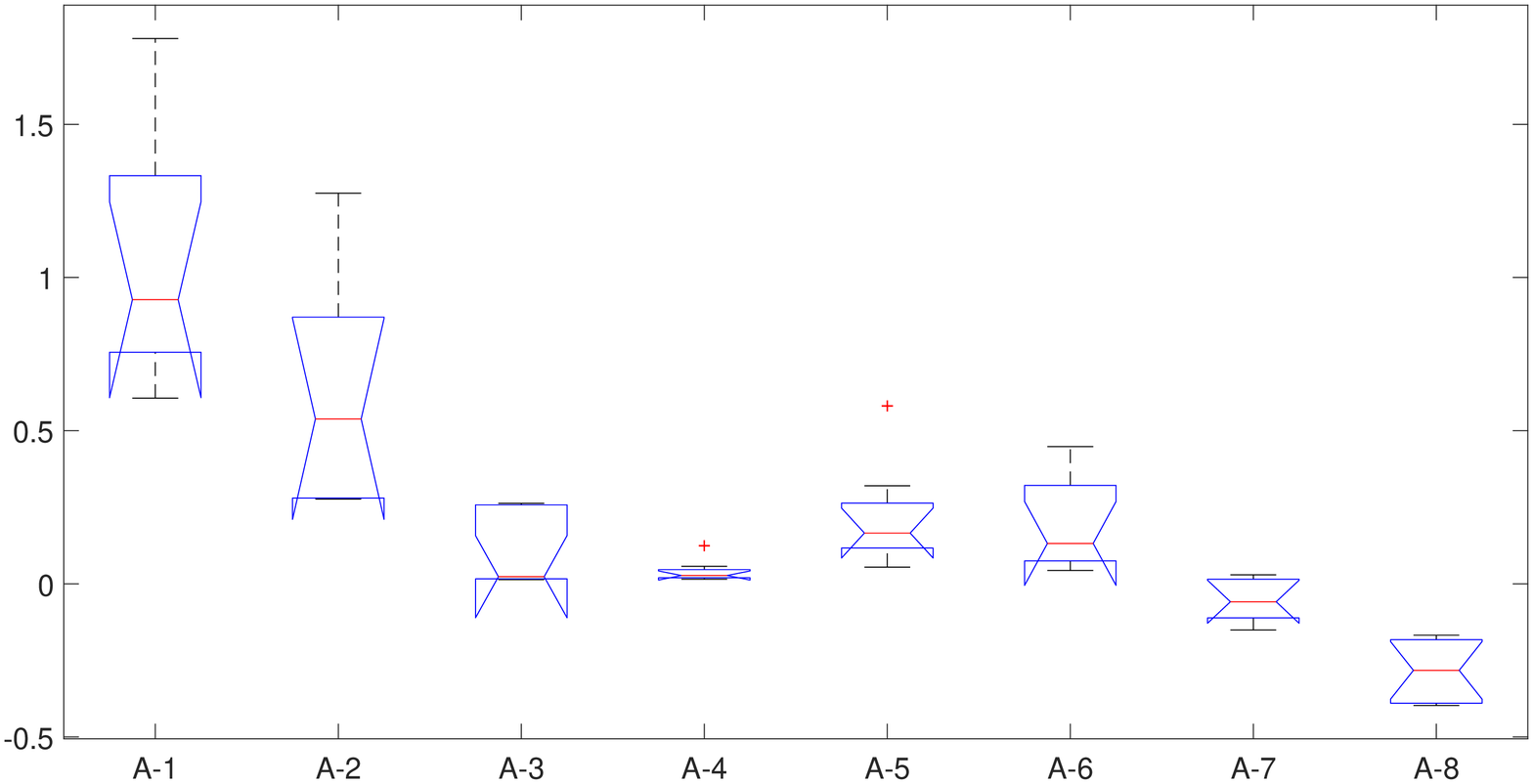}}\\
	\text{\bf a)}\\
	{\includegraphics[width=0.8\textwidth]{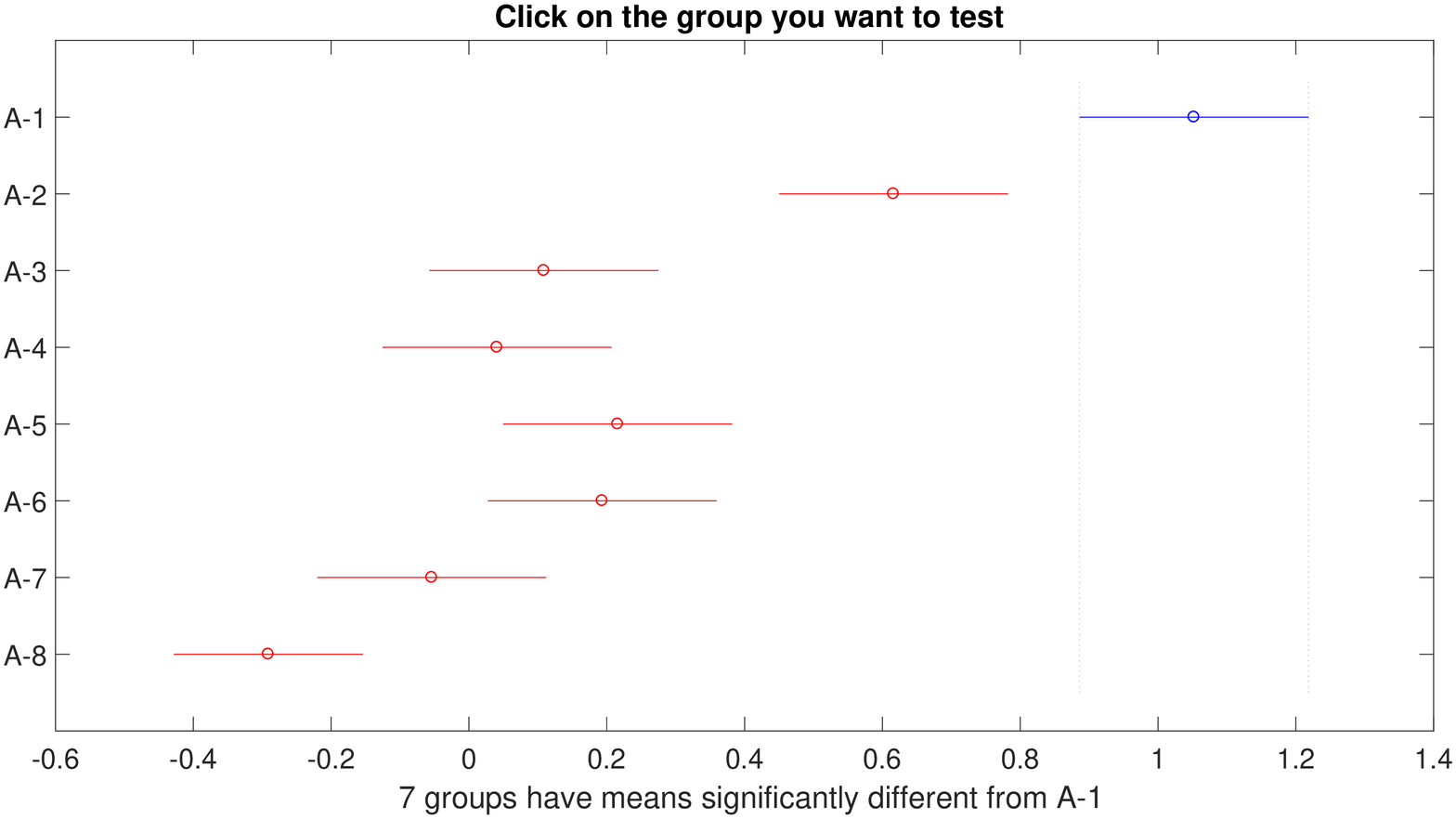}}\\
	\text{\bf b)}\\
	\caption{{\bf a).} Boxplot; {\bf b).} Multiple comparison test.} 
		\label{Fig:ANOVA}
\end{figure}

\begin{table}[ht!]
	\centering
	\caption{The ANOVA Table shows the between-groups (Groups) and the within-groups (Error) variation. ``SS" is the sum of squares and ``df"  means degrees of freedom associated to SS. MS indicates the mean squared error, i.e. the estimate of the error variance. The value of the F-statistic is given by the ratio of the mean squared errors.}
	\begin{tabular}{lcccccc}\\ \hline
		\bf Source&\bf SS&\bf df&\bf MS&\bf F&\bf Prob$>$F\\ \hline
		Groups & 10.8783 &  7 & 1.55405 & 34.71 & 8.00796e-19\\
		Error  &  2.6862 & 60 & 0.04477\\
		Total  & 13.5645 & 67\\
		\hline
	\end{tabular}
\label{tab:ANOVA}
\end{table}
Furthermore, the boxplot (Figure \ref{Fig:ANOVA}.a)) and a  multiple comparison test performed on the eight groups (Figure \ref{Fig:ANOVA}.b)) have suggested partitioning the data sample into  the following four groups of observations 1--8, \ 9--16, \ 17--56,\ 57--68.

\subsection{Jumps fitting by translation}\label{section3.2}
Since  the CIR model \eqref{CIRrate} does not fit negative interest rates and normal/high volatility when the rate value is small, spot rates must be shifted away from zero and/or negative values.

One chance is given by an affine type transformation, as follows
$$
r_{shift} (t) = \hat{\mu}_{r(t)} + \hat{\sigma}_{r(t)}\, r_{real}(t), \quad t\in [0,T]
$$
where $\hat{\mu}_{r(t)}$ and $\hat{\sigma}_{r(t)}$ are the sample mean and the sample standard deviation of $r_{real}(t),$ respectively. But, in practice, this is not the best choice due to some potential issues such as persistence of negative values, worse fitting, changes in the short interest rates behaviour, etc. Among different options, a translation of type
\begin{equation}\label{t1}
r_{shift}(t) = r_{real}(t) + \alpha,\quad t\in [0,T],
\end{equation}
with $\alpha>0$ a constant term, is a reasonable choice because it leaves unchanged the stochastic dynamics of short interest rates, i.e. $dr_{shift}(t)=dr_{real}(t)$. 
The value of the parameter $\alpha$ can be arbitrarily selected but, in our opinion, the most appropriate choice must take into account the empirical distribution of interest rates. For our purpose,  we set $\alpha$ equal to the approximate value, calculated from the available market data, corresponding to the $99th$-percentile of the conditional distribution of the process $\{r_{real}(t), t\in[0,T]\}.$ 
However, if further negative values are  between the $99th$- and the $100th$-percentile, then $\alpha$ can be set equal to the approximate value corresponding to the $1st$-percentile of the conditional distribution of spot rates. Hence the translation \eqref{t1} becomes $r_{shift}(t)=r_{real}(t)-{\alpha}$. 

The translation is applied after a check carried out on each group partitioning the original data sample; the check consists in calculating the harmonic mean - because it is more robust than the arithmetic one in the presence of extreme values - and in verifying whether it is smaller than a constant value  arbitrarily  chosen small  (e.g. $10^{-2}$). If this  happens in at least one group, then the whole sample is translated.

\subsection{Sub-optimal ARIMA models}\label{section3.3}
The second step consists in deriving the best fitting  ARIMA$(p,i,q)$ model to each group of interest rates partitioning  the original market data sample. Thus we start by selecting, for each group, a set of ARIMA $(p,i,q)$ models whose standardized residuals satisfy the following ``sub-optimal" conditions:
\begin{enumerate}
	\item Absence of both autocorrelation (AC) and partial autocorrelation (PAC) in the time series\footnote{If this condition is not verified, we can require just the absence of autocorrelation.};
	\item Absence of unit roots (stationarity of the time series);
	\item Normally distributed standardized residuals; 
	\item $R^{2}_{ARIMA}>0.5$,
\end{enumerate} 
where $R^{2}_{ARIMA}$ denotes the statistics $R^2$, defined in \eqref{rsquare}, computed for the ARIMA $(p,i,q)$ model. We look for only the indices $i\in\{0,1,2\}$ and $p,q\in\{1,2,3\}.$ 

As mentioned, to ensure that the residuals of the selected ARIMA $(p,i,q)$ models look like a Gaussian white noise, the Johnson's transformation (Johnson (1949)\cite{Johnson1949}) is applied to the standardized residuals. The Johnson's method consists in developing a flexible system of distributions, based on three families of transformations, that translates an observed, non-normal distribution to one conforming to the standard normal distribution. The transformation of a non-normal random variable $X$ to a standard normal variable $Z$ is written as
\begin{equation}\label{Jtransform}
Z=\gamma +\delta f\left(\frac{X-\xi}{\lambda}\right),\quad \lambda,\; \delta>0
\end{equation}
where $f$ is  function of a simple form. In particular, $f((X-\xi)/\lambda)$ must be a monotonic function of $X$, and its range of values have to correspond to the actual range of possible values of $(X-\xi)/\lambda.$ The parameters $\delta$ and $\gamma$ reflect respectively the skewness and kurtosis of $f$, while $\xi $ and $\lambda $ are the mean and the standard deviation of $X.$ The algorithm to estimate the four parameters $\gamma$, $\delta$, $\lambda$ and $\xi$, and perform the appropriate transformation is available as a Matlab Toolbox written by Jones (2014) \cite{Jones}).

In the sequel the normally distributed standardized ARIMA residuals applies to the random part of the CIR model to simulate trajectories of the interest rate process and calibrate the parameters $k, \theta, \sigma$ in \eqref{CIRrate} to the (eventually) translated market interest rates.

\subsubsection{Calibration of CIR parameters}\label{section3.3.1}
For the sake of simplicity, we rewrite the SDE  \eqref{CIRrate} as follows
\begin{align}\label{CIRQ}
dr(t) &= (\tilde{k}(\tilde{\theta} - r(t)) - \lambda r(t))\, dt + \sigma\sqrt{r(t)}\, dW(t) \nonumber \\
&= (k(\theta - r(t))\, dt + \sigma\sqrt{r(t)}\, dW(t),
\end{align} 
and we set
$$
k=\tilde{k} + \lambda >0, \quad \theta=\frac{\tilde{k}\tilde{\theta}}{k}>0.
$$

Consider the  $j{th}$-group partitioning the available market data sample, which we assume to be of length $n_j$. The calibration of the CIR parameters in the group is performed as follows
\begin{enumerate}
	\item The volatility $\sigma$ is estimated by the group standard deviation, namely $\hat{\sigma}_j$;
	\item The long-run mean parameter $\theta$ is estimated by the group mean, namely $\hat{\theta}_j$;
	\item The speed of mean reversion $k$ is estimated by that value, say $\hat{k}_j,$ solving the following minimization problem:
	\begin{equation}\label{7}
	\min\limits_{k>0}\, S_{j}(k) := \min\limits_{k>0}\, \sqrt{\frac{\sum\limits_{h=1}^{n_j} (u^{j}_{h}(k) - \overline{u}^j (k))^{2} }{n_j-1}}.
	\end{equation}
\end{enumerate}  
For any $k>0,$ we define
\begin{equation}\label{8}
u^{j}_{h} (k) := r^{j}_{h}(k) - r^{j}_{shift, h}, \quad h=1,\cdots,n_j,
\end{equation}
being $r^{j}_{shift, h}$ the real shifted interest rate value, and  $r^{j}_{h}(k)$ the corresponding simulated CIR interest rate value expressed as a function of the unknown parameter $k$. The $r^{j}_{h}(k)$ are calculated by applying the strong convergent Milstein discretization scheme (1979) \cite{Milnstein} to the SDE \eqref{CIRQ}. 
Brigo \& Mercurio (2006) \cite[Section 22.7]{BrigoMercurio2006} showed that the Milstein scheme converges in a much better way than other numerical schemes for the CIR process. It reads as
\begin{equation}\label{9}
r^j_{h+1}(k) = r^j_h (k) + k(\hat{\theta}_j - r^j_h)\, \Delta + \hat{\sigma}_j\sqrt{r^j_h \Delta}\; Z^j_{h+1} + \frac{(\hat{\sigma}_j)^2}{4}\, [(\sqrt{\Delta}\; Z^j_{h+1})^{2} - \Delta],
\end{equation}
where  $\Delta$ is the time step - we set $\Delta = 1/30$ due to monthly observed data - and $Z^{j}_{h+1}$ are standard normally distributed random variables.  Indeed, in this case, $Z^{j}_{h+1}$ are  the normally distributed standardized residuals of each ARIMA $(p,i,q)$ model selected for the $jth$-group. They are computed by applying the Johnson's transformation \eqref{Jtransform}, with the ARIMA residuals as realizations of the random variable $X$. 

After calculation of the estimates $(\hat{k}_j,\, \hat{\theta}_j,\, \hat{\sigma}_j),$ the CIR fitted values to the shifted observed spot rates in the $jth$-group, $r^{j}_{shift, h}$, are computed by the simulation scheme \eqref{9} as follows
\begin{equation}\label{CIRfitted}
\hat{r}^j_{h+1} = \hat{r}^j_{h} + \hat{k}_j(\hat{\theta}_j - \hat{r}^j_{h})\, \Delta + \hat{\sigma}_j\sqrt{\hat{r}^j_{h}\Delta}\, Z^j_{h+1} + \frac{(\hat{\sigma}_j)^2}{4}\, [(\sqrt{\Delta}\, Z^j_{h+1})^{2} - \Delta],
\end{equation}
where  $\Delta$ and $Z^{j}_{h+1}$ are as before. To measure the goodness-of-fit, the statistics $R^{2}$ is computed. For sake of clarity, in the sequel we will denote by $R^{2}_{CIR}$ the statistics \eqref{rsquare} when referring to the CIR model.

\subsection{Optimal ARIMA-CIR model }\label{section3.4}
For each group $j$, the ``optimal" ARIMA$(p,i,q)$ model providing the best CIR fitting to real data will be chosen among the selected sub-optimal ARIMA $(p,i,q)$ models, as described in Section \ref{section3.3}, that satisfy the following additional conditions:
\begin{enumerate}
\item[5.] The ARIMA $(p,i,q)$ minimizes the Bayesian Information Criterion (BIC) matrix whose rows and columns are the possible $p$ and $q$ lags, respectively ({\em $BIC$ condition});
\item[6.] $R^{2}_{CIR}>0.5$. 
\end{enumerate}

Therefore we define the following sets of candidate ARIMA models:
$$
\mathcal{I}_{AC} = \left\lbrace (p,i,q)\, | \, \text{ARIMA$(p,i,q)$ satisfies conditions 1.-- 4. and 6.} \right\rbrace 
$$
and
$$
\mathcal{I}_{ACB} = \left\lbrace (p,i,q) |\, \text{ARIMA$(p,i,q)$ satisfies conditions 1.-- 6.} \right\rbrace. 
$$
Obviously, $\mathcal{I}_{ACB}\subset \mathcal{I}_{AC}$. 

Last but not least, the ``optimal" ARIMA model is chosen in the above defined classes as the model minimizing the  RMSE $\varepsilon_j$ 
\begin{equation}
\min\limits_{\hat{r}^j}\, \varepsilon_j := \min\limits_{\hat{r}^j}\; \sqrt{\frac{1}{n^j}\sum_{h=1}^{n^j}(r^j_{shift,h} - \hat{r}^j_h)^{2}}, 
\end{equation}
where the minimum is computed with respect to all the CIR fitted values vectors, $\hat{r}^j,$ simulated for the $jth$-group. 

The algorithm proposed in Table \ref{tab:ARIMA-CIR} finds, for each group $j,$  the ``optimal" ARIMA $(p,i,q)$ model and returns as output: the matrix of indices $(p,i,q)$ belonging to the sets $\mathcal{I}_{AC}$ and  $\mathcal{I}_{ACB},$ the corresponding CIR fitted values vector $\hat{r}^j$ computed by  \eqref{CIRfitted}, and the associated values of the statistics $R^2_{CIR}$ and  $\varepsilon_j$. The main steps of the algorithm can be summarized as follows:
\begin{table}[ht!]\caption{ARIMA-CIR algorithm}
	\begin{tabular}{|l|}
		\hline
		\textbf{Step 1:} verify if $check1=1$ for the $j{th}$-group;\\
		\textbf{Step 2:} if $check1=1$ verify if $check2=1$ for the current group;\\
		\textbf{Step 3:} if $check2=1$ print the output. Else, reduce the size $n^{j}$ of the current group\\
		 to $(n^{j}-m)$ where $m=8.$\\ 
		\textbf{Step 4:} repeat \textbf{Step 1}-\textbf{Step 3} for the remaining observations in the current group.\\ 
		\textbf{Step 5:} return to \textbf{Step 1} for the group $j+1.$\\
		\hline
	\end{tabular}
	\label{tab:ARIMA-CIR}
\end{table}

Note that $check1$ and $check2$ refer respectively to conditions 1.-- 5. and 6. Their value is equal to 1 if those conditions are satisfied. It is worth mentioning that a test on the efficiency of the above algorithm could be done by verifying that $\varepsilon^j$ is small for all $j$ and that the weighted mean of the $\varepsilon^j$ (see formula (\ref{total_error})) is small for the whole data sample. 

We applied the ARIMA-CIR algorithm to the $n=68$ monthly observed EUR interest rates with 1 day ( overnight) maturity, mentioned in Section \ref{sec:Dataset}. We recall that the ANOVA analysis suggested to partition the data sample into four groups of observations: 1--8, \ 9--16, \ 17--56,\ 57--68. Table \ref{tab:outputs} shows in detail the outputs for this sample. The group containing the observations 17--56 has been futher segmented into three sub-groups of size $m=8$ or a multiple thereof: 17--32, \ 33--48 and 49--56. The triplets $(p,i,q)$ identified by a rectangle in Table \ref{tab:outputs}, indicates the ``optimal" ARIMA model chosen for each group/sub-group (with the bigger $(R^{2}_{CIR})_j$ and the smaller $\varepsilon_j$ values). As it can be seen, none of these models fulfils  the $BIC$ \emph{condition}. 

\begin{table}[!htbp]\centering\caption{Outputs from the ARIMA-CIR algorithm for the 68 monthly EUR interest rates on overnight maturity}\label{tab:outputs}
\begin{threeparttable}\small
	\begin{tabular}{lcccccccc}\\ \hline
		\bf j&\bf group/sub-groups&\bf ARIMA model&\pmb{$(R^{2}_{CIR})_j$}&\pmb{$\varepsilon_{j}$}&\textbf{BIC cond.}\\ \hline
		1&1--8&\fbox{(2,0,1)}&0.8166&\fbox{0.1643}\tnote{1}\\
		&&(2,0,2)&0.5930&0.2414\\
		&&(3,0,2)&0.780&0.1865\\
		&&(1,1,1)&0.8166&0.1643\\
		&&(1,2,1)&0.7309&0.2026\\
		&&(1,2,2)&0.7805&0.1845&$\surd$\\
		&&(3,2,1)&0.7023&0.2104\\ \hline
		2&9--16&(1,0,1)&0.6842&0.2090\\
		&&(2,0,2)&0.7799&0.2588\\
		&&(3,0,2)&0.6418&0.2661&$\surd$\\
		&&(1,1,1)&0.7378&0.2043\\
		&&\fbox{(1,1,2)}&0.8472&\fbox{0.1554}\\
		&&(2,1,1)&0.6842&0.2169\\
		&&(2,1,2)&0.7799&0.2012\\ 
		&&(3,1,2)&0.6418&0.2333&\\\hline
		3&17--32&\fbox{(1,0,3)}&0.9174&\fbox{0.0326}\\
		&&(3,0,1)&0.5485&0.0646\\ \hline
		4&33--48&\fbox{(3,0,1)}&0.6901&\fbox{0.0833}\\ 
		&&(3,1,2)&0.6332&0.1146\\
		&&(3,2,1)&0.6332& 0.1146\\ \hline
	\end{tabular}
	\end{threeparttable}
\end{table}

\begin{figure} [!h]
\includegraphics[width=1\textwidth]{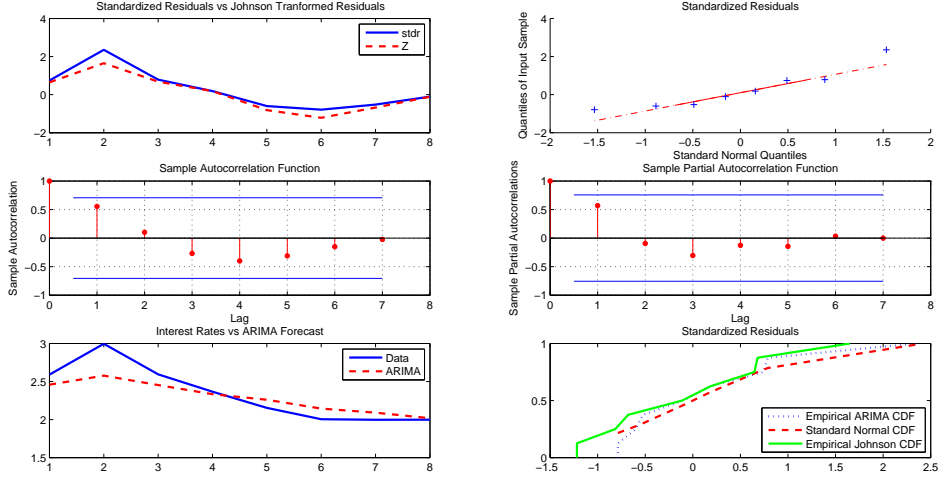}
\caption{Qualitative statistical analysis related to the group 9--16. {\bf Top line:} ARIMA $(1,1,2)$ standardized residuals versus Johnson's transformed residuals ({\em left}); Q-Q normal plot for the ARIMA $(1,1,2)$ standardized residuals ({\em right}). {\bf Middle line:} AC plot ({\em left}) and PAC plot ({\em right}). {\bf Bottom line:} real interest rates versus ARIMA $(1,1,2)$ fitted values ({\em left}); comparison among the cumulative distribution function (CDF) of the standard normal distribution, the empirical CDF of ARIMA $(1,1,2)$ standardized residuals and the empirical CDF of the residuals after the Johnson's transformation ({\em right}).}
\label{fig:subgroup}
\end{figure}

\begin{table}[!htbp]\centering
\begin{threeparttable}\small
	\begin{tabular}{lcccccccc}\\ \hline	
	\bf j&\bf group/sub-groups&\bf ARIMA model&\pmb{$(R^{2}_{CIR})_j$}&\pmb{$\varepsilon_{j}$}&\textbf{BIC cond.}\\ \hline
	5&49--56&(1,0,1)&0.5076&0.0597\\
	&&(1,0,2)&0.6030&0.0526\\
	&&(2,0,1)&0.5702&0.0577\\
	&&(3,0,1)&0.7648&0.0483\\
	&&(3,0,2)&0.6240&0.0537\\
	&&(1,1,1)&0.5702&0.0577\\
	&&(2,1,1)&0.5393&0.0588\\
	&&\fbox{(3,1,2)}&0.7715&\fbox{0.0479}\\
	&&(1,2,1)&0.5542&0.0582\\
	&&(3,2,1)&0.6987&0.0479\\
	&&(3,2,2)&0.6343&0.0536&$\surd$\\ \hline	
        6&57--68&(1,0,2)&0.5964&0.0752\\
		&&(1,0,3)&0.8080&0.0570\\
		&&(2,0,2)&0.8136&0.0559\\
		&&(3,0,2)&0.7899&0.0577\\
		&&(3,0,3)&0.6560&0.0704\\
		&&(1,1,2)&0.8136&0.0559\\
		&&(1,1,3)&0.8006&0.0614\\
		&&\fbox{(2,1,1)}&0.8239&\fbox{0.0486}\\
		&&(2,1,2)&0.8542&0.0525\\
		&&(2,1,3)&0.8936&0.0551\\
		&&(3,1,2)&0.9023&0.0511\\
		&&(3,1,3)&0.8000&0.0580\\ \hline		
		\end{tabular}
		
		\begin{tablenotes}
    \item[1] For a more exact comparison we use the numeric format long.
  \end{tablenotes}
\end{threeparttable}
\end{table}
Figure \ref{fig:subgroup} below reports the qualitative statistical analysis carried out by applying the ARIMA $(1,1,2)$ model chosen as the ``optimal" forecasting model for the group 9--16 (similar plots for the other group/sub-groups are reported in  \ref{A-MainGraphs}). 

\begin{figure}[!htbp]
		\centering
		\includegraphics[width=0.9\textwidth]{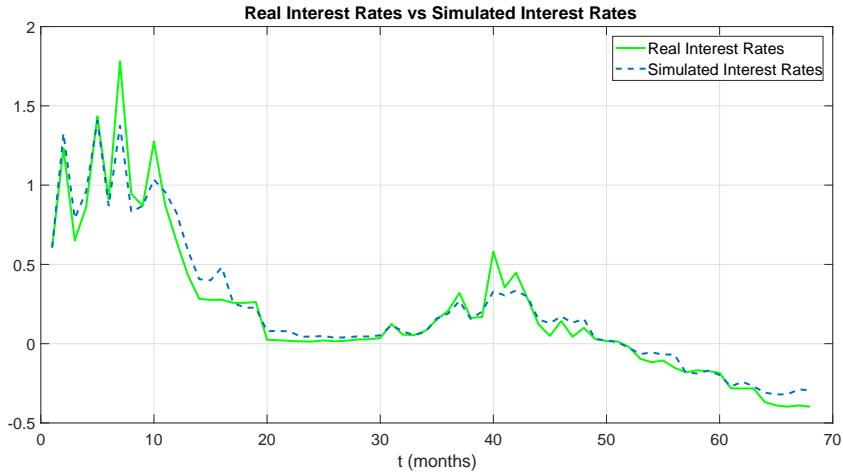}
	\caption{Monthly EUR interest rates with T=1 day (overnight) maturity vs CIR fitted rates}
	\label{fig:CIRfittedvalues1}
\end{figure}

Figure \ref{fig:CIRfittedvalues1} compares the short-term interest rates structure of the analysed market data sample with the corresponding curve of CIR fitted values computed by the simulation scheme \eqref{CIRfitted} for each group partitioning the whole data sample. It is worth noting that the normally distributed standardized residuals of the ``optimal" ARIMA-CIR model selected for each group/sub-group (see Table \ref{tab:outputs}) replace the realizations of a standard Brownian motion in \eqref{CIRfitted}. This strategy allows us  to get an exact trajectory of CIR fitted values instead of a curve averaged over 100,000  simulated trajectories. Consequently, the computational cost is considerably reduced. 

The real interest rates have been shifted by using a translation of type \eqref{t1}, where $\alpha$ corresponds to the $99^{th}$-percentile of the conditional distribution of the spot rates process, as described in Section \ref{section3.2}. Finally, the CIR fitted values have been shifted back.

The total values of the statistics $R^{2}_{CIR}$ and the RMSE $\varepsilon$ have been computed on the whole sample as a weighted mean of the $(R^{2}_{CIR})_j$ and $\varepsilon_{j}$ values corresponding to the ``optimal" ARIMA model chosen for each group/sub-group, i.e.
$$
\widetilde{R}^{2}_{CIR}= \sum_{j=1}^{J} \frac{n_j}{n}(R^{2}_{CIR})_j,
$$
\medskip
\begin{equation}\label{total_error}
\varepsilon= \sqrt{\sum_{j=1}^{J} \frac{n_j}{n}\sum_{h=1}^{n_j} (r^j_{shift,h} - \hat{r}^j_h)^{2}}.
\end{equation}

The Appendix \ref{B-Estimates} reports the CIR parameters estimates and the plots of the function $S_j(k),$ defined in \eqref{7}, for all groups/sub-groups.

We would like to emphasize that the data sample segmentation does not imply a change in the bond's pricing formula \eqref{CIRprice}, but only affects the estimation of the CIR parameters $k,\theta, \sigma$ in \eqref{CIRfitted} that are locally calibrated to each group/sub-group of data.

\subsection{The change points detection problem}
\noindent As explained in Section \ref{section3.1}, where we have partitioned the available market data sample in groups with an arbitrary fixed length, say $m=8$, the main difficulty concerns the choice  of the optimal segmentation to detect abrupt changes in the variance of the interest rates dynamics. In the literature there exist several approaches for detecting multiple changes  in the probability distribution of a stochastic process or a time series such as sequential analysis (i.e., ``online" methods), clustering based on maximum likelihood estimation (i.e. ``offline" methods), minimax change detection, etc. (see, for example, Bai and Perron (2003) \cite{Bai}, Lavielle (2005) \cite{Lavielle2005} and (2006) \cite{Lavielle2006}, Hacker and Hatemi-J (2006)\cite{Hacker}, Adams and MacKay (2007) \cite{Adams}, Arlot and Cenisse (2011) \citep{Arlot}). 

\pagebreak
\begin{table}[!ht]
\small
	\caption{Numerical scheme for change points detection by Lavielle's algorithm}
	\label{T-11}
	\centering
	\begin{threeparttable}
		\begin{tabular}{|l|}
			\hline
			- compute  $v(1:end)$ the array of change points detected in the real data array $x$\\
			\  \ by the Lavielle method\\
			- set $l=v(1)$;\\
			- initialize $xstart=1$, $xend=l$; \\
			\  \ ($xstart, \, xend$ denote the first and last component of a partitioning group\\
			\  \  at each processing cycle)\\
			- initialize $j=1$;\\
			- set $smax=xstart$ +1 (each group must have a minimum length equal to $2$)\tnote{1};\\
			- \textbf{while} $l<v(end)$ \& $l\not=smax-1$ \\ 
			- compute $check \ 1$ and the matrix $L$\\ ($L$ is the matrix of possible ARIMA $(p,i,q)$ for $x(xstart:xend)$);\\
			- \textbf{if} $check1=1$\\
			- compute $check2$; \\
			- \textbf{if} $check2=1$\\
			- compute $\varepsilon_{j}$ and $R^{2}_{CIR};$\\ 
			- let $ex(j)=l;$ \\
			\ \ ($ex$ is the array of the rescaled change points, see Figure \ref{fig:pointsdetection} );\\ 
			- set $xstart=ex(j)+1$;\\
			- set $j=j+1$;\\
			- set $l=v(j)$;\\
			- set $xend=l$;\\
			- \textbf{else} \\
			- set $l=l-1;$\\
			- set $xend=l$;\\
			- \textbf{end}\\
			- \textbf{end}\\
			- \textbf{if} $l=smax$\\
			- \textbf{if} $length(v)\leq j+1$ \\
			- set $l=v(j+1)$;\\
			- \textbf{else} $break$; \\
			- \textbf{end}\\
			- \textbf{end}\\
			- \textbf{end}\\
			\hline
		\end{tabular}
	\begin{tablenotes}
    \item[1] Some statistical tests (involved in $check1$) require a minimum sample length equal to $7$, so one can also set $smax=xstart+6$.
  \end{tablenotes}
\end{threeparttable}
\end{table}

\begin{figure}[h]
	\centering
	\includegraphics[width=1.0\textwidth]{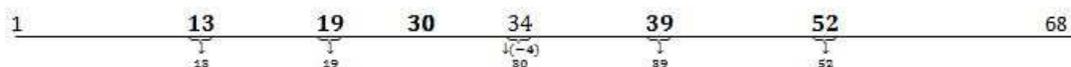}
	\caption{Scheme for change points detection from the algorithm in Table \ref{T-11}}
	\label{fig:pointsdetection}
\end{figure}

\clearpage
In this work we decided to implement the Matlab algorithm proposed by Lavielle (2005) \cite{Lavielle2005} for the detection of changes in the variance, which led to partitioning the real data sample into the following six groups: 1--13, 14--19, 20--30, 31--39, 40--52, 53--68. However, taking into account that our CIR\# model is based on the combination of an ARIMA model and the original CIR model, the above detected change points reported in Figure \ref{fig:pointsdetection}, namely 13, 19, 34, 39, 52, have to be adjusted according to the numerical scheme described in Table \ref{T-11}.

\begin{figure}[!htbp]
	\centering
	\includegraphics[width=0.90\textwidth]{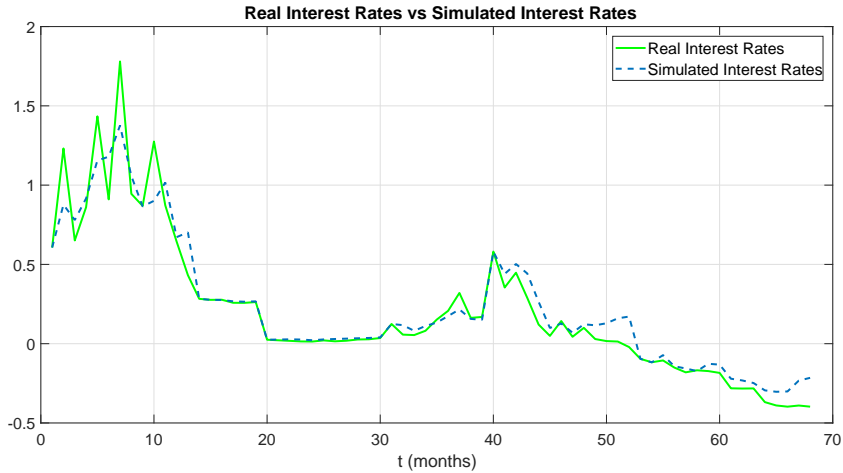}
	\caption{Monthly EUR interest rates with T=1 day (overnight) maturity vs CIR\#  fitted rates (after application of the Lavielle method)}
	\label{fig:CIRfittedvalues2}
\end{figure}

Table \ref{tab:outputs1} lists the results from the ARIMA-CIR algorithm after application of the change point detection algorithm. Figure \ref{fig:CIRfittedvalues2} plots the real interest rates versus the CIR\#  fitted values according to results in Table \ref{tab:outputs1}. We found that the total $R^{2}_{CIR}=0.7584$ and the total RMSE $\varepsilon=0.4159$. As before, for all $j$, the errors $\varepsilon_{j}$ are at most of the order of $10^{-1}$. 

\begin{table}[!htbp]\small
	\centering
	\caption{Outputs from the ARIMA-CIR algorithm applied to $n=68$ monthly EUR interest rates with T=1 day (overnight) maturity (after application of the Lavielle method)}
	\begin{tabular}{lccccccc}\\ \hline
		\bf j&\bf Groups&\bf ARIMA model&\pmb{${(R^{2})}^j_{CIR}$}&\pmb{$\varepsilon^{j}$}&\textbf{BIC cond.}\\ \hline
		1&1--13&\fbox{(2,1,1)}&0.6223&\fbox{0.2251}\\
        &&(3,1,1)&0.6117&0.2258\\		
		&&(3,1,2)&0.5985&0.2299\\ \hline
		2&14--19&(1,0,1)&0.8842&0.0048\\
        &&(2,0,1)&0.7960&0.0062\\
		&&(2,0,2)&0.8814&0.0047\\
		&&\fbox{(2,0,3)}&0.8795&\fbox{0.0047}\\		
		&&(1,1,1)&0.8291&0.0058\\
		&&(1,1,2)&0.8821&0.0048\\
		&&(1,2,1)&0.7623&0.0068\\
		&&(2,1,1)&0.8421&0.0055\\  \hline
	 \end{tabular}
\label{tab:outputs1}
\end{table}
\begin{table}[!htbp]\centering
	\begin{tabular}{lccccccc}\\ \hline
		\bf j&\bf Groups&\bf ARIMA model&\pmb{${(R^{2})}^j_{CIR}$}&\pmb{$\varepsilon^{j}$}&\textbf{BIC cond.}\\ \hline
		3&20--30&(3,1,2)&0.6345&0.0087&$\surd$\\
		&&(3,1,3)&0.6193&0.0089\\ 
		&&(1,2,3)&0.6452&0.0099\\ 
		&&(2,2,2)&0.5178&0.0096&$\surd$\\
		&&(2,2,3)&0.5777&0.0089\\  
		&&\fbox{(3,2,2)}&0.6369&\fbox{0.0085}\\ \hline
		4&31--39&(1,0,1)&0.7165&0.0444\\
		&&(1,0,2)&0.6895&0.0465\\
		&&(1,0,3)&0.6815&0.0469\\
		&&(1,1,1)&0.7168&0.0439\\
		&&(1,1,2)&0.7247&0.0415\\
		&&(1,1,3)&0.5409&0.0546\\
		&&(2,1,1)&0.6876&0.0458\\
		&&\fbox{(2,1,2)}&0.7478&\fbox{0.0404}\\
		&&(2,1,3)&0.6778&0.0449\\
		&&(1,2,3)&0.5911&0.0507\\
		&&(2,2,3)&0.5817&0.0509\\
		&&(3,2,2)&0.6379&0.0473\\ \hline
		5&40--52&\fbox{(1,0,2)}&0.8841&\fbox{0.1050}\\
		&&(2,0,1)&0.8868&0.1196\\
		&&(2,0,2)&0.7985&0.1317&$\surd$\\
		&&(2,0,3)&0.7979&0.1315\\ 
		&&(3,0,1)&0.8944&0.1178\\
		&&(3,0,2)&0.8186&0.1230\\
		&&(3,0,3)&0.8786&0.1158\\
		&&(1,1,1)&0.6359&0.1653\\
		&&(1,1,2)&0.6004&0.1752&$\surd$\\
		&&(1,1,3)&0.6436&0.1595\\
		&&(2,1,1)&0.8459&0.1287\\
		&&(2,1,2)&0.5547&0.1783\\
		&&(2,2,1)&0.6694&0.1637\\
		&&(3,2,3)&0.7306&0.1630\\ \hline
		6&53--68&\fbox{(1,0,3)}&0.8111&\fbox{0.0683}\\ \hline
	\end{tabular}
\end{table}

\section{Forecast of future interest rates}
\begin{figure}[!ht]
		\centering
		\includegraphics[width=.90\textwidth]{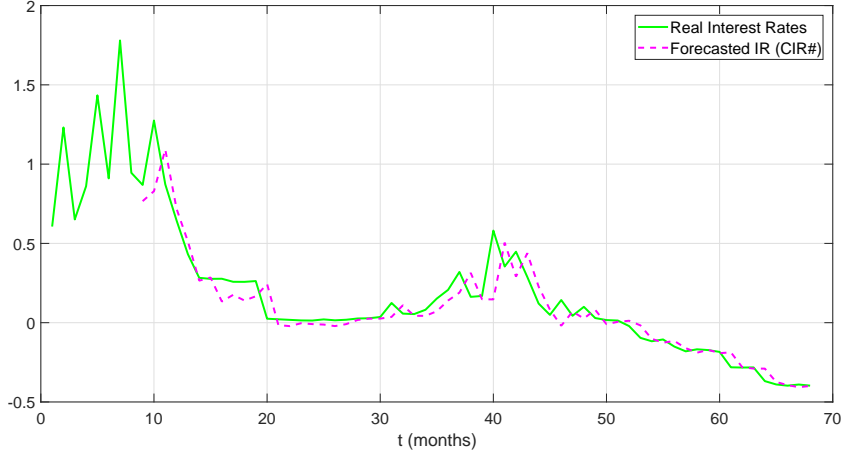}
	\caption{{\bf Forecast of the next-month interest rate based on a rolling window of 8 real data:}  monthly  EUR interest rates with maturity T=1 day (overnight) versus predicted next-month interest rates.}
\label{Fig:vsc2}	
\end{figure}

In this section we will address briefly the CIR\# model's progress on future interest rate forecasts from a window of observed market data. Ex-ante forecasts require a thorough analysis and will be extensively treated in a forthcoming research.
It is worth noting that in this work we decided to impose the most challenging conditions by modelling the shortest part of the yield curve (e.g. the overnight rate) and using only a handful of number of observations. For instance, with monthly data we have found that $m=8$ observations are sufficient for a good calibration. Thus we start to consider a fixed size window of 8 real interest rates that is
rolled through time, each month adding the new rate and taking off the oldest rate. The length of this window (8 months) is the historical period over which we forecast the next-month spot rate value. The numerical procedure described in Sections \ref{section3.2}--\ref{section3.4} has been applied to forecast future next-month interest rates based on monthly EUR interest rates with overnight maturity. The predicted curve is shown in Figure \ref{Fig:vsc2}, compared with the real observed term structure.

It is evident that the predicted next-month spot rates computed by the CIR\# model follow the market trend. Moreover, the values of $R^2_{CIR}$ and RMSE $\varepsilon,$ computed to measure the goodness-of-fit of forecast interest rates to real data, are respectively $0.8045$ and $0.1392$. 

\subsection{CIR\# forecasts versus CIR forecasts}
\begin{figure}[!h]
	\centering
	\includegraphics[width=.90\textwidth]{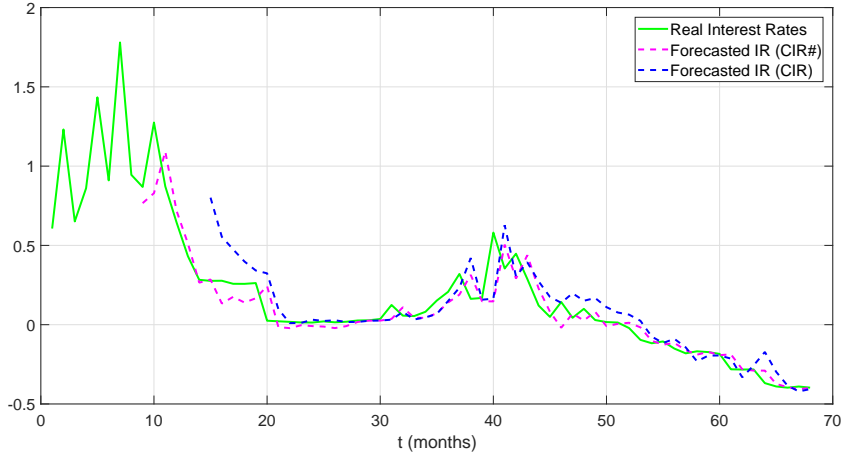}
	\caption{{ \bf Forecast of future next-month expected interest rates:} monthly EUR interest rates with overnight maturity compared with  future next-month interest rates predicted by the CIR\# model based on a rolling window of $m=8$ real data, and future next-month interest rates predicted by the classical CIR model based on a rolling window of $m=14$ real data.}
	\label{Fig:vs2}	
\end{figure}

Here we are going to analyze the CIR\# improvements in forecast as compared to the original CIR model.  It is worth noting that calibrating the CIR parameters $(k,\theta,\sigma)$  to real data, a hystorical window of $m>8$ observations is usually needed.  To do that, herein we applied  the martingale estimating functions method for diffusion processes proposed by Bibby et al. \cite[Example 5.4]{Bibby}(2005), which shows a better performance with respect to the Maximum Likelihood approach, as empirically confirmed  in \cite{OMB2}. In this case, to forecast a next-month rate by the standard CIR model the minimum length of a rolling window is $m \ge 14$, against the window size of $m=8$ observations, required by the CIR\#. Figure \ref{Fig:vs2} illustrates the future next-month values predicted by the CIR\# compared with the ones forecasted by the classical $CIR$ model, showing a better fitting to the real observed term structure.

%
The values of $R^2$ and  $\varepsilon$, listed in Tables \ref{tab:ForRsquare} and \ref{tab:ForRMSE} for ten term structures ranging from maturity $T= 1$ day to $T=270$ days, confirm the improvement of the proposed novel model compared to the original $CIR$ one applied to shifted data. 

\begin{table}[!h]
	\small
	\centering\spacingset{1}
	\caption{CIR\# versus CIR forecasts of future next-months interest rates for ten different maturities. Each column reports the $R^2$ values for both the models.}
	\label{tab:ForRsquare}
	\begin{tabular}{c|cccc} 
		\hline\noalign{\smallskip}\spacingset{2}
		 & \bf A & \bf B & \bf C & \bf D \\
		\noalign{\smallskip}\hline\noalign{\smallskip}
		\bf Maturity & $R^2$ & $R^2$ & Difference & Performance\\
         &  CIR\# & CIR & \bf A-B & \bf C/A \\
     	\noalign{\smallskip}\hline\noalign{\smallskip}\spacingset{3}
1 d  & 0.8082 & 0.6279 & 0.1803 & 22.38\%  \\
30 d & 0.9506 & 0.9232 & 0.0274 & 2.88\% \\
60 d & 0.9676 & 0.9234 & 0.0445 & 4.59\% \\
90 d & 0.9674 & 0.9343 & 0.0332 & 3.43\% \\
120 d  & 0.9705 & 0.9090 & 0.0615 & 6.33\%\\
150 d & 0.9735 & 0.9533 & 0.0202 & 2.07\%\\
180 d  & 0.9752 & 0.9567 & 0.0185 & 1.89\%\\
210 d & 0.9758 & 0.9588 & 0.0170 & 1.74\% \\
240 d & 0.9782 & 0.9589 & 0.0193 & 1.97\% \\
270 d & 0.9735 & 0.9589 & 0.0146 & 1.49\% \\
	\noalign{\smallskip}\hline 
\end{tabular}
\end{table}

\bigskip
\begin{table}[!h]
	\small
	\centering\spacingset{1}
	\caption{CIR\# versus CIR forecasts of future next-months interest rates for ten different maturities. Each column reports the RMSE $\varepsilon$ values for both the models.}
	\label{tab:ForRMSE}
	\begin{tabular}{c|cccc} 
		\hline\noalign{\smallskip}\spacingset{2}
		& \bf A & \bf B & \bf C & \bf D \\
		\noalign{\smallskip}\hline\noalign{\smallskip}
		\bf Maturity & $\varepsilon$ & $\varepsilon$ & Difference & Performance\\
		&  CIR\# & CIR & \bf A-B & \bf C/A  \\
		\noalign{\smallskip}\hline\noalign{\smallskip}\spacingset{3}
		1 d  & 0.0963 & 0.1416 & -0.0453  & 47.04\% \\
		30 d & 0.0463 & 0.0576 & -0.0113  & 24.40\% \\
		60 d & 0.0406 & 0.0628 & -0.0222 & 54.67\% \\
		90 d & 0.0476 & 0.0675 & -0.0199 & 41.80\% \\
		120 d  & 0.0484 & 0.0851 & -0.0367  & 75.82\% \\
		150 d & 0.0488 & 0.0646 & -0.0158  & 32.37\%\\
		180 d  & 0.0500 & 0.0660 & -0.0160  & 32.00\% \\
		210 d & 0.0511 & 0.0667 & -0.0157  & 30.72\% \\
		240 d & 0.0503 & 0.0689 & -0.0186 & 36.97\% \\
		270 d & 0.0570 & 0.0709 & -0.0139 & 24.38\%  \\
		\noalign{\smallskip}\hline 
	\end{tabular}
\end{table}

\section{Conclusions}
Several different extensions of the original model have been proposed to date, with the aim of overcoming the limitations of the CIR model: from one-factor models including time-varying coefficients or jump diffusions to multi-factor models. All these extensions preserve the positivity of interest rates but, in some cases, the analytical tractability of the basic model is violated. Our approach, instead, is based on a proper translation of interest rates such that the market volatility structure is maintained as well as the analytical tractability of the original CIR model. Thus the suggested \emph{CIR\# model} is quite powerful for the following reasons.
First, all the improvements are obtained within the CIR framework in order to preserve the single-factor property and the analytical tractability of the original model. Second, market interest rates are properly translated  away from zero and/or negative values. The market data sample is partitioned into sub-groups in order to capture all the statistically significant changes of variance in real spot rates and therefore gives an account of jumps. Third, we have introduced a new way of calibration of the CIR model parameters to actual data. The standard Brownian motion process in the random part of the model is replaced with normally distributed standardized residuals of ``optimal" ARIMA models
suitably chosen. As a result, exact CIR fitted values to real data are calculated and the computational cost of the numerical procedure is considerably reduced. Fourth, we have shown that the \emph{CIR\# model} is efficient and able to follow very closely the structure of market short-term interest rates  (especially for short maturities that, notoriously, are very difficult to handle) and to predict future interest rates better then the original CIR model. As a measure of goodness-of-fit, we obtained high values of the statistics $R^{2}$ and small values of the square error $\varepsilon$ for each sub-group and the entire data sample.   
Future research will show the predictive power of the model by extending the dataset in terms of frequency and size.
%



\clearpage
\begin{appendix}
\setcounter{figure}{0}
\setcounter{table}{0}
\section{Qualitative analysis related to Table \ref{tab:outputs}} \label{A-MainGraphs} 

We report the qualitative statistical analysis carried out for each group/sub group according to the results reported in  Table \ref{tab:outputs}. The qualitative analysis related to the results in Table \ref{tab:outputs1} is analogous and so it is omitted.
\begin{figure}[h] 
	\centering
	\includegraphics[width=1\textwidth]{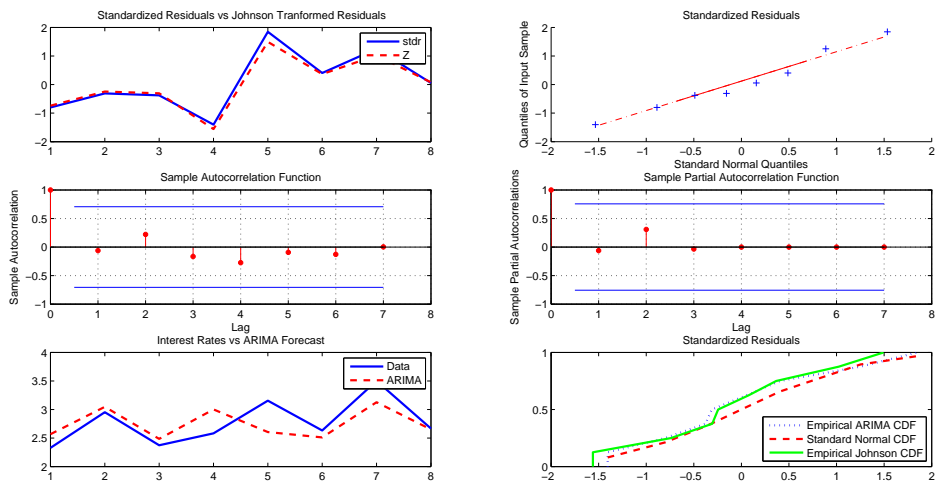}
\caption{\small Qualitative statistical analysis related to the sub-group 1--8. {\bf Top line:} ARIMA $(2,0,1)$ standardized residuals versus Johnson's transformed residuals ({\em left}); Q-Q normal plot for the ARIMA $(2,0,1)$ standardized residuals ({\em right}). {\bf Middle line:} AC ({\em left}) and PAC ({\em right}) plots. {\bf Bottom line:} real interest rates versus ARIMA $(2,0,1)$ fitted values ({\em left}); comparison of the standard normal  cumulative distribution function (CDF) with the empirical CDF of ARIMA $(2,0,1)$ standardized residuals and of Johnson's transformed residuals ({\em right}).}
\end{figure}
\begin{figure}[!htbp]
	\centering
	\includegraphics[width=1\textwidth]{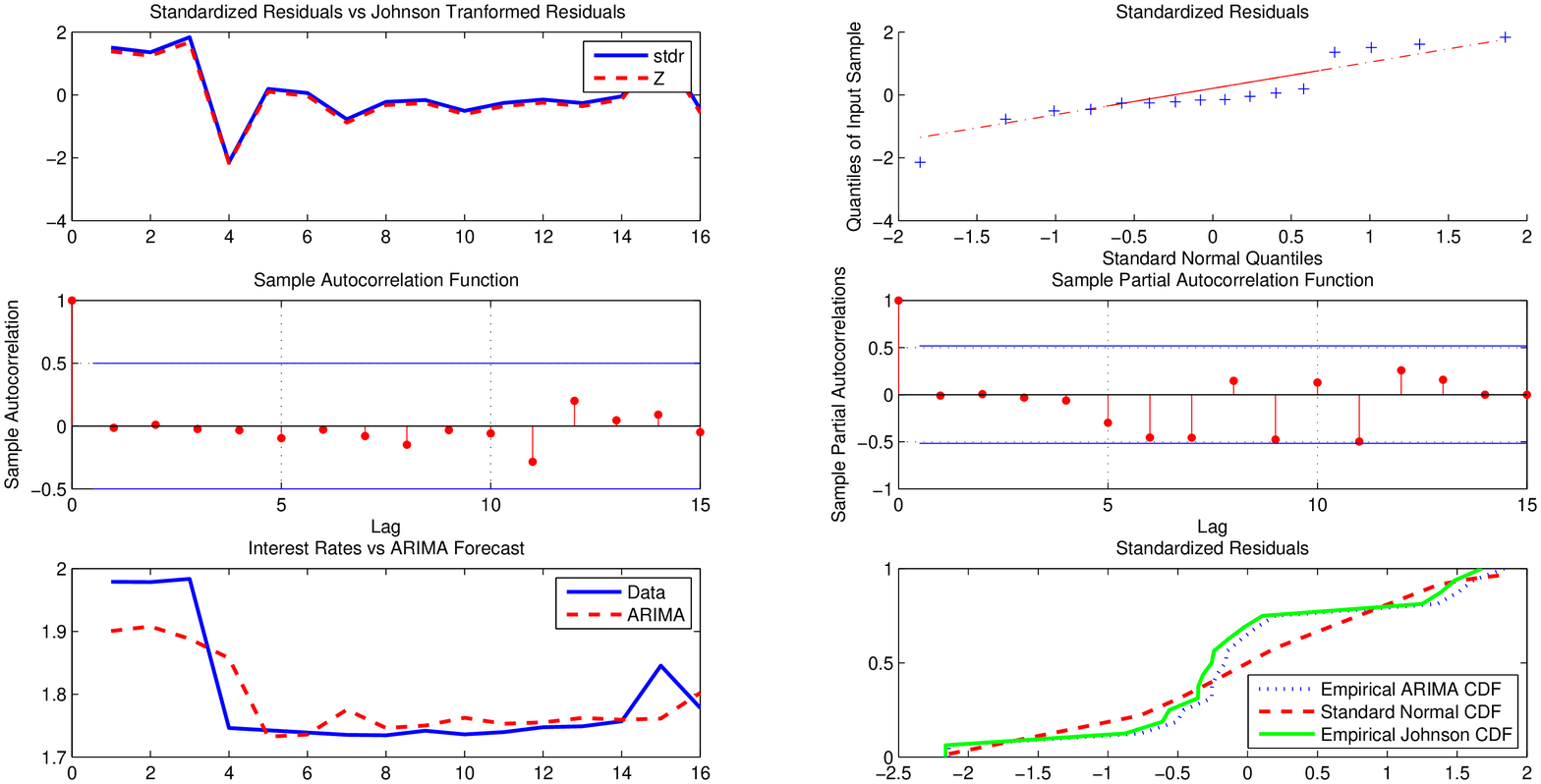}
\caption{\small Qualitative statistical analysis related to the sub-group 17--32. {\bf Top line:} ARIMA $(1,0,3)$ standardized residuals versus Johnson's transformed residuals ({\em left}); Q-Q normal plot for the ARIMA $(1,0,3)$ standardized residuals ({\em right}). {\bf Middle line:} AC ({\em left}) and PAC ({\em right}) plots. {\bf Bottom line:} real interest rates versus ARIMA $(1,0,3)$ fitted values ({\em left}); comparison of the standard normal cumulative distribution function (CDF) with the empirical CDF of ARIMA $(1,0,3)$ standardized residuals and of Johnson's transformed residuals ({\em right}).}
\end{figure}
\begin{figure}[!htbp]
	\centering
	\includegraphics[width=1\textwidth]{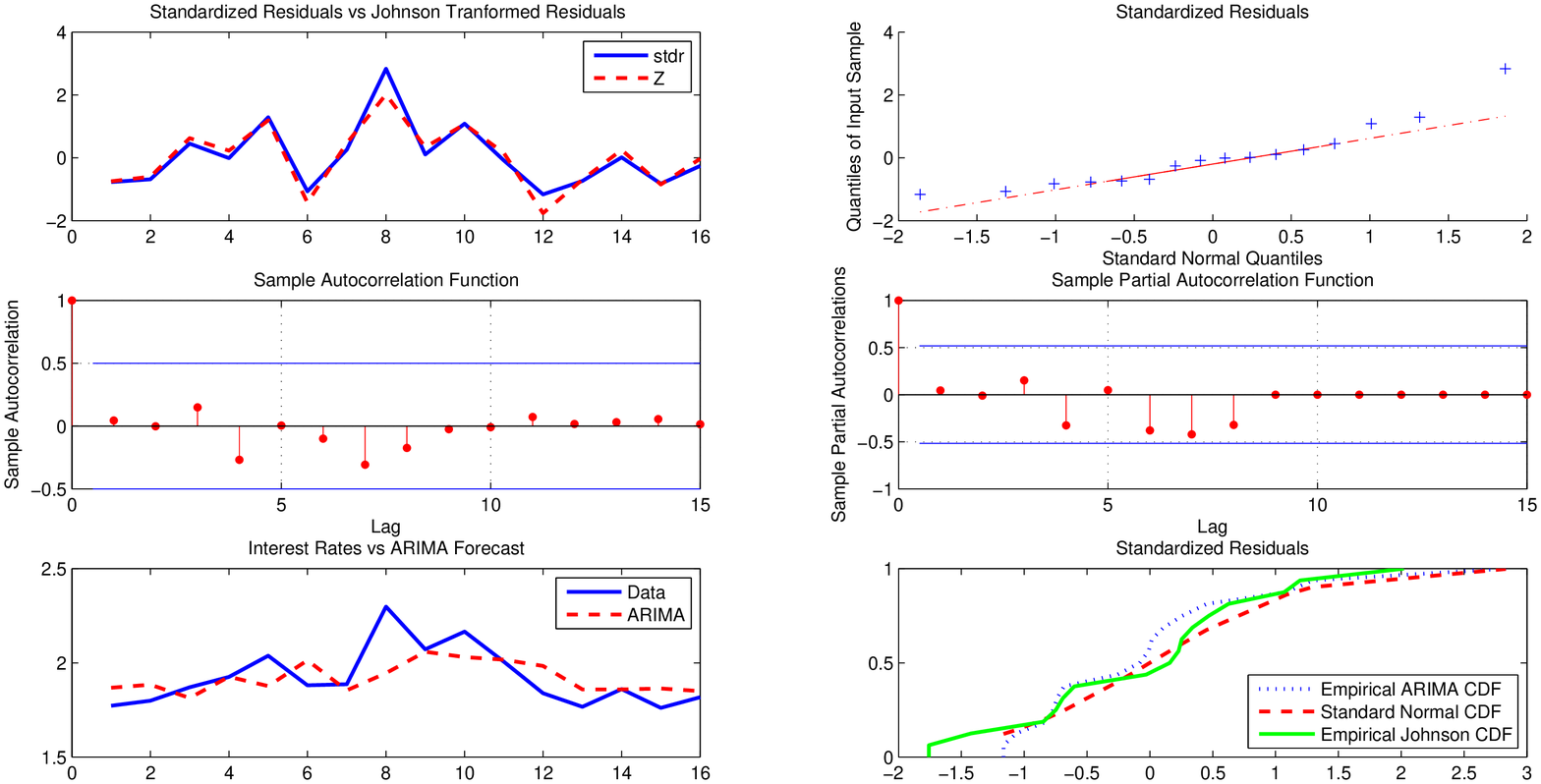}
	\caption{\small Qualitative statistical analysis related to the sub-group 33--48. {\bf Top line:} ARIMA $(3,0,1)$ standardized residuals versus Johnson's transformed residuals ({\em left}); Q-Q normal plot for the ARIMA $(3,0,1)$ standardized residuals ({\em right}). {\bf Middle line:} AC ({\em left}) and PAC ({\em right}) plots. {\bf Bottom line:} real interest rates versus ARIMA $(3,0,1)$ fitted values ({\em left}); comparison of the standard normal  cumulative distribution function (CDF) with the empirical CDF of ARIMA $(3,0,1)$ standardized residuals and of Johnson's transformed residuals ({\em right}).}
\end{figure}
\begin{figure}[!htbp]
	\centering
	\includegraphics[width=1\textwidth]{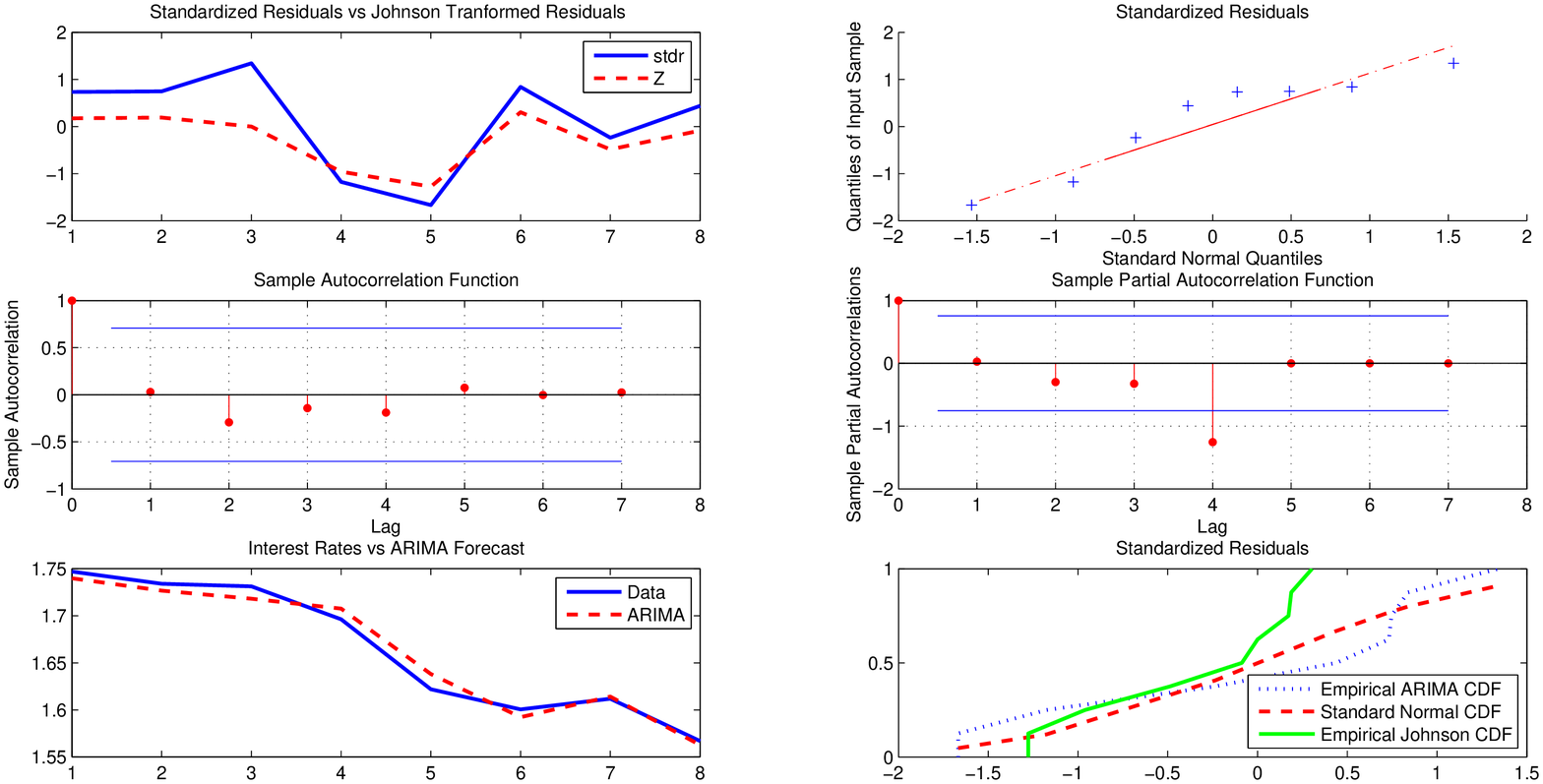}
	\caption{\small Qualitative statistical analysis related to the sub-group 49--56. {\bf Top line:} ARIMA $(3,1,2)$ standardized residuals versus Johnson's transformed residuals ({\em left}); Q-Q normal plot for the ARIMA $(3,1,2)$ standardized residuals ({\em right}). {\bf Middle line:} AC ({\em left}) and PAC ({\em right}) plots. {\bf Bottom line:} real interest rates versus ARIMA $(3,1,2)$ fitted values ({\em left}); comparison of the standard normal  cumulative function (CDF) with the empirical CDF of ARIMA $(3,1,2)$ standardized residuals and of Johnson's transformed residuals ({\em right}).}
\end{figure}
\begin{figure}[!htbp]
	\centering
	\includegraphics[width=1\textwidth]{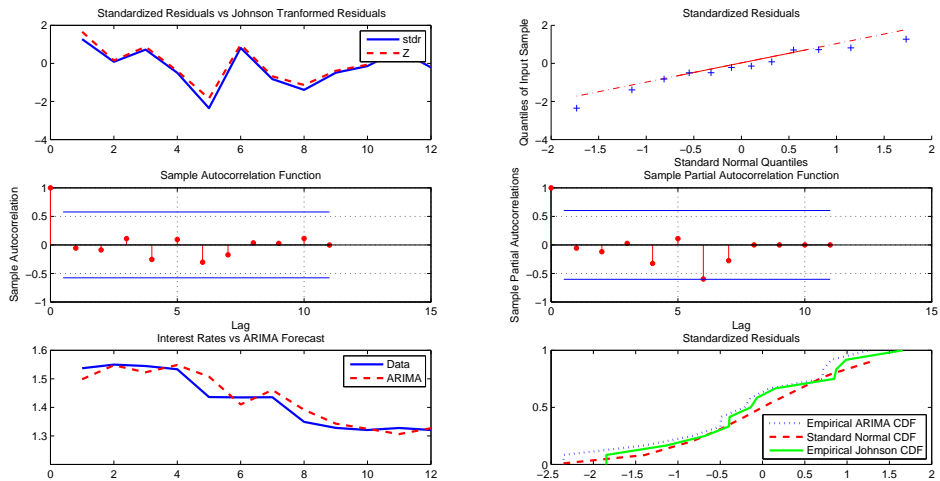}
\caption{\small Qualitative statistical analysis related to the sub-group 57--68. {\bf Top line:} ARIMA $(2,1,1)$ standardized residuals versus Johnson's transformed residuals ({\em left}); Q-Q normal plot for the ARIMA $(2,1,1)$ standardized residuals ({\em right}). {\bf Middle line:} AC ({\em left}) and PAC ({\em right}) plots. {\bf Bottom line:} real interest rates versus ARIMA $(2,1,1)$ fitted values ({\em left}); comparison of the standard normal cumulative distribution function (CDF) with the empirical CDF of ARIMA $(2,1,1)$ standardized residuals and of Johnson's transformed residuals ({\em right}).}
\end{figure}

\par
\clearpage

\section{CIR parameter estimates}\label{B-Estimates}
\noindent We report the  estimates of the CIR parameters $k,\theta,\sigma$  and, in particular, the plots of the function $S_j(k)$ defined in \eqref{7}, corresponding to the selected  ``optimal" ARIMA models reported in Table \ref{tab:outputs}. 
\begin{figure}[h]
	\centering
	\includegraphics[width=1\textwidth]{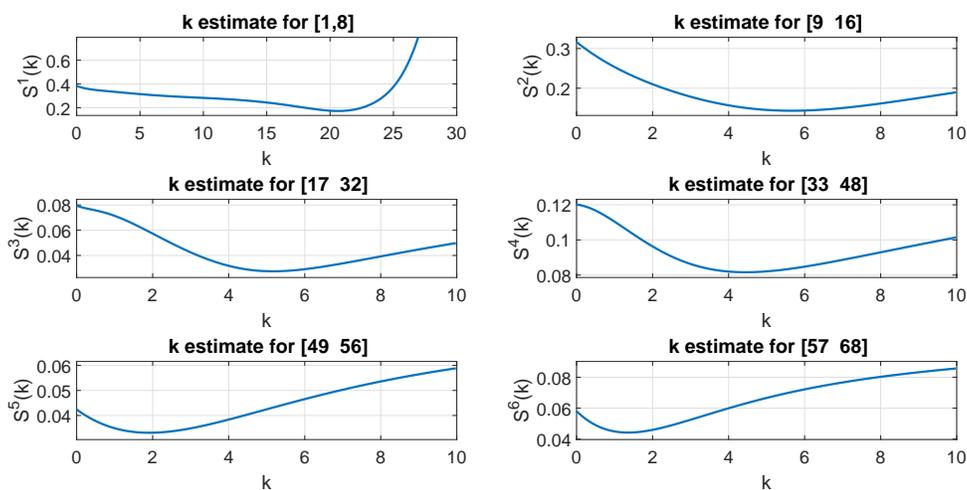}
	\caption{\small Plots of the functions $S_j(k)$ for each group/sub-group}
\end{figure}

\begin{table}[!htbp]\centering\caption{CIR parameter estimates based on 68 monthly observed 1-day (overnight) EUR interest rates}
	\begin{tabular}{ccrrr}\\ \hline 
		\bf j&\bf group/sub-group&\pmb{$\hat{k}_j$} &\pmb{$\hat{\theta}_j$}&\pmb{$\hat{\sigma}_j$}\\ \hline \medskip
		1&1--8&20.6364&2.7699&0.4027\\ \medskip
		2&9--16&5.6621&2.3338&0.3663\\  \medskip
		3&17--32&5.1649&1.7924&0.0954\\  \medskip
		4&33--48&4.4462&1.9223&0.1546\\  \medskip
		5&49--56&1.9092&1.6637&0.0709\\  \medskip
		6&57--68&1.3555&1.4264&0.0958\\  \hline
	\end{tabular}
	\label{tab:estimates}
\end{table}
\end{appendix}

\end{document}